\documentclass[10pt]{iopart}

\usepackage{graphicx}

\begin{document}

\title{Quadrupolar correlations and deformation effect on two neutrino $\varepsilon
\beta ^{+}$ and $\varepsilon \varepsilon $ modes of $^{156}$Dy isotope}
\author{P K Rath$^{1}$, R Chandra$^{1,2}$, S Singh$^{1}$, P K Raina$^{2}$
and J G Hirsch$^{3}$}
\address{
$^{1}$Department of Physics, University of Lucknow, Lucknow-226007, India\\
$^{2}$Department of Physics and Meteorology, IIT, Kharagpur-721302, India\\
$^{3}$Instituto de Ciencias Nucleares, Universidad Nacional Aut\'{o}noma de
M\'{e}xico, A.P. 70-543, M\'{e}xico 04510 D.F., M\'{e}xico}

\begin{abstract}
The two-neutrino positron double-$\beta $ decay modes of $^{156}$Dy isotope
are studied in the Projected Hartree-Fock-Bogoliubov framework for the $%
0^{+}\rightarrow 0^{+}$ transition. Theoretically calculated half-lives of 
electron-positron conversion and double-electron capture modes are
presented. The effect of
the quadrupolar
 deformation on nuclear transition matrix element $M_{2\nu }$ 
is also investigated.
\end{abstract}


\maketitle

\section{Introduction}

The nuclear $\beta \beta $ decay, which is  mediated by strangeness conserving 
charged weak current is expected to occur through double-electron emission 
$\left(\beta ^{-}\beta ^{-}\right) $, double-positron emission $\left( \beta
^{+}\beta ^{+}\right) $, electron-positron conversion $\left( \varepsilon
\beta ^{+}\right) $ and double-electron capture $\left( \varepsilon
\varepsilon \right) $ with the emission of two neutrinos, no
neutrinos, single Majoron and double Majorons. The $\beta
^{+}\beta ^{+} $$/$$\varepsilon 
\beta ^{+}$$/$$\varepsilon\varepsilon$ processes are 
energetically competing and we
refer to them as $e^{+}\beta \beta $ decay. The $\left( \beta \beta \right)
_{2\nu }$ decay conserves the lepton number exactly and is an allowed
process in standard model of electroweak 
interactions
($SM$). The lepton number violating $\left( \beta \beta \right) _{0\nu }$ decay, which is
theoretically possible in many a gauge 
models beyond the $SM$, has not
been observed so far. All the present experimental efforts are devoted to
its observation, which would immediately imply that neutrinos are massive
Majorana particles. In comparison to the $\beta ^{-}\beta ^{-}$ decay, less
attention has been paid to study the $e^{+}\beta \beta $ decay due to
relatively low Q-values and low abundances of $e^{+}\beta \beta $ emitters.
The experimental and theoretical developments in the study of nuclear $%
e^{+}\beta \beta $ decay have been excellently reviewed over the past years \cite
{rose65,verg83,verg02,doi92,doi93,bara95,suho98,kirp00,klapDBD,bara04}.

The half-life $T_{1/2}^{2\nu }$ of $\left( \beta \beta \right) _{2\nu }$
decay is a product of accurately known phase space factor $G_{2\nu }$ and
nuclear transition matrix element (NTME) $M_{2\nu }$. Hence, the validity of
different models employed for nuclear structure calculations can be tested
by comparing the calculated NTMEs $M_{2\nu }$ with those extracted from the
experimentally observed half-lives $T_{1/2}^{2\nu }$ for the 
$0^{+}\rightarrow 0^{+}$ transition of $\left( \beta ^{-}\beta
^{-}\right) _{2\nu }$ mode. In contrast to the $\left( \beta ^{-}\beta
^{-}\right) _{2\nu }$ mode, which has been observed in ten $\beta ^{-}\beta
^{-}$ emitters out of 35 potential candidates, the $\left( e^{+}\beta \beta
\right) _{2\nu }$ decay modes  
have not been observed experimentally so far. However,
limits on their half-lives $T_{1/2}^{2\nu }$ have been given for 24 out of
34 possible isotopes \cite{tret02}. Meshik \textit{et al} 
have reported on a positive
observation of $^{130}$Ba decay with half-life $T_{1/2}=(2.16\pm 0.52)\times
10^{21}$ y for all modes \cite{mesh01} and this result is
consistent with theoretical expectations for the $\left( \varepsilon \varepsilon
\right) _{2\nu }$ mode \cite{hirs94}. This value is in slight
contradiction with the
experimental
 limit $T_{1/2}>4\times 10^{21}$ y 
reported in
\cite{bara96}. If
confirmed, it would be the very first observation of $e^{+}\beta \beta $
decay. In the absence of experimental data, there is no 
way to
judge the reliability of present nuclear structure calculations. In
principle, the $\beta ^{-}\beta ^{-}$ decay and $e^{+}\beta \beta $ decay
can provide the same information. The observation of $\left( e^{+}\beta
\beta \right) _{2\nu }$ decay will be interesting from the nuclear structure
point of view, as it will be a challenging task to calculate the NTMEs of
these modes along with the $\left( \beta ^{-}\beta ^{-}\right) _{2\nu }$
mode in the same theoretical framework and it will further constrain the
nuclear models employed to study the $\beta \beta $ decay. In
addition, the observation of $\left( e^{+}\beta \beta \right) _{0\nu }$
decay will be helpful in deciding finer issues like dominance of mass
mechanism or admixture of the right handed current in the electroweak interaction 
\cite{hirs94}.

Rosen and Primakoff were the first to study the $\left( e^{+}\beta \beta
\right) _{2\nu }$ decay theoretically \cite{rose65}. Later on, Kim and
Kubodera estimated the half-lives of all the three modes with estimated
NTME and non-relativistic phase space factors \cite{kim83}. Abad 
\textit{et al} performed similar calculations using relativistic Coulomb
wave functions \cite{abad84}. In addition, some other theoretical studies 
were done by Zel'dovich \textit{et al} \cite{zeld81}, 
Eramzhyan \textit{et al} \cite{eram82}, Bernabeu \textit{et al} 
\cite{bern83} and Balaev \textit{et al} \cite{bala89}.
The $\left( e^{+}\beta \beta \right) _{2\nu }$
decay has been studied mainly in three types of models, namely shell model
and its variants, the quasiparticle random phase approximation (QRPA) and
its extensions and alternative models \cite{suho98}. The nuclear many-body
problem is solved as exactly as possible in the shell-model and it is the
best choice for the calculation of the NTMEs. However, the number of basis
states increases quite drastically for most of the $\beta \beta $ decay
emitters as they are medium or heavy mass nuclei and hence, Vergados has
studied the $\left( e^{+}\beta \beta \right) _{2\nu }$ decay of $^{58}$Ni, $%
^{92}$Mo and $^{96}$Ru nuclei in the weak coupling limit. Over the past
years, the large scale shell-model calculations have been successfully
performed to study the potential $\beta \beta $ emitters \cite
{caur96,enge96,radh96,skou86}. In the shell-model, the $\left( e^{+}\beta
\beta \right) _{2\nu }$ decay modes of $^{40}$Ca \cite{chin84}, 
$^{36}$Ar, $^{54}$Fe, $^{58}$Ni \cite{naka96} and $^{92}$Mo \cite{suho97} 
isotopes have been investigated.

The QRPA has emerged as the most successful model in explaining the
observed quenching of NTMEs by incorporating the particle-particle part of the
effective nucleon-nucleon interaction in the proton-neutron channel \cite
{voge86,civi87} and the observed half-lives $T_{1/2}^{2\nu }$ of several $\left( \beta
\beta \right) _{2\nu }$ decay emitters were reproduced successfully \cite
{suho98}. In the QRPA model, Staudt \textit{et al} have evaluated the NTMEs 
for $\left(\beta ^{+}\beta ^{+}\right) _{2\nu }$ mode \cite{stau91}. The $\left(
e^{+}\beta \beta \right) _{2\nu }$ decay modes were studied in the QRPA and its
extensions \cite{hirs94,bara96,suho93,auno96,toiv97,suho01,stoi03}. In spite of 
the 
success of the QRPA in the study of $\beta \beta $ decay, there
is a need to include the deformation degrees of freedom in its formalism.
The deformed QRPA model has been developed for studying $\beta \beta $ decay
of spherical as well as deformed nuclei. However, these studies have been mainly
restricted to $\beta ^{-}\beta ^{-}$ decay so far \cite{pace04,alva04,rodi08}. 
Besides the QRPA, the other models employed to study the $\left( e^{+}\beta
\beta \right) _{2\nu }$ decay modes are the  SU(4)$%
_{\sigma \tau }$ \cite{rumy98}, SSDH \cite{civi98} and pseudo-SU(3) \cite{cero99}.

In fact, the subtle interplay of pairing and
quadrupolar correlations present in the effective two-body interaction
decides the complex structure of nuclei. In addition to the pairing interaction,
which  plays an important role in all even $Z$-even $N$ $%
\beta \beta $ emitters, the crucial role of deformation
degrees of freedom in the structure of $^{100}$Mo and $^{150}$Nd
isotopes has been already reported \cite{grif92,suho94}. In the 
Projected Hartree-Fock-Bogoliubov (PHFB) model, the two crucial components 
of effective two body interaction, namely pairing and 
quadrupolar correlations are
incorporated on equal footing and the rotational symmetry is restored by
projection technique providing wave functions with good angular momentum for
the parent and daughter nuclei involved in the $\beta \beta $ decay. However,
the PHFB model is unable to provide information about the structure of 
intermediate odd $Z$-odd $N$ nuclei in its present version and hence, on the
single-$\beta $ decay rates and the distribution of GT strength. In spite of
this limitation, the PHFB model, in conjunction with pairing plus
quadrupole-quadrupole ($PQQ$) interaction \cite{bara68}, has been
successfully applied to study the $0^{+}\rightarrow 0^{+}$ transition
of $\left( \beta ^{\pm }\beta ^{\pm }\right)_{2\nu }$ modes
\cite{chan05,shuk05,rain06,sing07}, where it was
possible to describe the lowest excited states of the parent and daughter
nuclei along with their electromagnetic transition strengths, as well as to
reproduce their measured $\beta ^{-}\beta ^{-}$ decay rates \cite
{chan05,sing07}.

In the PHFB model, the existence of an inverse correlation
between the quadrupole deformations and the magnitudes of NTMEs $M_{2\nu }$ has
been shown \cite{chan05,rain06,sing07}. In addition, it has been
observed that the NTMEs are usually large in the absence of quadrupolar
correlations. With the inclusion of the quadrupolar correlations, the NTMEs
are almost constant for small admixture of the $QQ$ interaction and
suppressed substantially in realistic situation. For similar deformations of 
parent and daughter nuclei, the
NTMEs have well defined maximum \cite{chan09}. In the Interacting Shell
Model (ISM) \cite{caur08,mene08}, similar observations have been
also reported. Presently, we aim to study the $e^{+}\beta \beta $ decay 
of $^{156}$Dy isotope. 
The deformation parameters 
$\beta_2$ for parent $^{156}$Dy and daughter $^{156}$Gd nuclei are 
$0.2929\pm0.0016$ and $0.3376\pm0.0018$, respectively, which will provide 
a typical case study of the deformation effect. The present paper is 
organised as follows. We briefly present in section 2 the required 
theoretical formalism. In section 3, we calculate half-lives 
$T_{1/2}^{2\nu} $ for the $0^{+}\rightarrow 0^{+}$ transition of 
$\left( \varepsilon \beta ^{+}\right) _{2\nu }$ and $\left(
\varepsilon \varepsilon \right) _{2\nu }$ modes for $^{156}$Dy isotope
together with various
spectroscopic properties, specifically, yrast spectra, 
reduced $B(E2$:$0^{+}\rightarrow 2^{+})$ 
transition probabilities, quadrupole moments $Q(2^{+})$
and gyromagnetic factors $g(2^{+})$ of $^{156}$Dy and $^{156}$Gd nuclei. The
expressions to calculate these spectroscopic properties in the PHFB model
are given in \cite{dixi02}. Moreover, we study the effect of
deformation on NTME $M_{2\nu }$ of $\left( e^{+}\beta \beta \right) _{2\nu }$
decay modes vis-a-vis the changing strength of the $QQ$ interaction. Finally, 
the concluding remarks are presented in section 4.

\section{Theoretical framework}

The inverse half-lives for the $0^{+}\rightarrow 0^{+}$ transition 
of $\left( e^{+}\beta \beta \right) _{2\nu }$
decay modes are given by 
\begin{equation}
\left[ T_{1/2}^{2\nu }(\beta )\right] ^{-1}=G_{2\nu }(\beta )\left| M_{2\nu
}\right| ^{2}
\end{equation}
where $\beta $ denotes $\left( \beta
^{+}\beta ^{+}\right)_{2\nu} $$/$$\left( \varepsilon
\beta ^{+}\right)_{2\nu} $$/$$\left( \varepsilon\varepsilon \right)_{2\nu} $ mode. 
The phase space factors $\ G_{2\nu }(\beta )$\ have been calculated
to good accuracy \cite{doi92} and the nuclear model dependent NTME $M_{2\nu
} $ is expressed as

\begin{equation}
M_{2\nu }=\sum\limits_{N}\frac{\langle 0_{F}^{+}||\mathbf{\sigma }\tau
^{-}||1_{N}^{+}\rangle \langle 1_{N}^{+}||\mathbf{\sigma }\tau
^{-}||0^{+}\rangle }{(E_{N}-E_{I})+E_{0}}.  \label{m2n}
\end{equation}
To evaluate equation (\ref{m2n}), one has to explicitly sum over all states
of the intermediate odd $Z$-odd $N$ nuclei. However, it is not possible to study
the structure of intermediate odd-odd nuclei in the present version of the
PHFB model. Alternatively, the summation over the intermediate states can be
carried out by using the summation method \cite{civi93}, when the GT
operator commutes with the effective two-body interaction \cite
{cast94,jghi95,cero99}. We use the latter procedure to sum over the
intermediate states for evaluating the NTME $M_{2\nu }$ \cite
{chan05,rain06,sing07}.

Presently, the nuclear wave functions are generated in the HFB framework by
using a Hamiltonian consisting of an effective two-body interaction with
pairing and quadrupolar correlations \cite{bara68}. Explicitly, the Hamiltonian 
is written as 
\begin{equation}
H=H_{sp}+V(P)+\zeta _{qq}V(QQ)
\end{equation}
\smallskip \noindent where $H_{sp},$ $V(P)$ and $V(QQ)$ denote the single
particle Hamiltonian, the pairing and $QQ$ part of the effective two-body
interaction. Further, $\zeta _{qq}$ denotes the strength of $QQ$ part of the
effective two-body interaction. The purpose of introducing it is to study
the role of deformation by varying the strength parameter $\zeta _{qq}$. The
final results are obtained by setting the $\zeta _{qq}$ = 1.

The two-body part of the Hamiltonian given by Eq. (3) contains only pairing 
and quadrupole-quadrupole interactions.
The 
quadrupole-quadrupole interaction commutes with the
Gamow-Teller operator (a complete proof can be found in Ref. [54]).
In principle, we should employ a pairing interaction, which includes not
only the proton-proton and neutron-neutron T=1 channels, but also
proton-neutron T=1 and T=0 channels. With them a pairing interaction can be
built, which commutes with the GT operator [56] when both channels
have the same strengths. On the other side, it has been
shown that both the T=0 and T=1 proton-neutron
gaps go to zero [57] at the mean field level for medium to heavy mass
nuclei with $N-Z > 6$. While it would be very important to
include a spin-isospin channel in the Hamiltonian, this can be done
beyond mean field, for example, by employing the QRPA over the PHFB model.
For the restricted mean field calculations presented below, there is
no practical difference in employing the full pairing interaction or 
pairing only in the proton-proton and neutron-neutron channels.
For simplicity, we use the latter.
It follows that the NTME $M_{2\nu }$ for the 
0$^{+}\rightarrow 0^{+}$ transition of 
$\left(e^{+}\beta \beta \right) _{2\nu }$ decay in the PHFB model in
conjunction with the summation method is given by \cite{rain06,sing07}

\begin{eqnarray}
M_{2\nu } &=&\sum\limits_{\pi ,\nu }\frac{\langle {\Psi _{00}^{J_{f}=0}}%
||\left( \mathbf{\sigma .\sigma }\tau ^{-}\tau ^{-}\right) _{\pi \nu }||{%
\Psi _{00}^{J_{i}=0}}\rangle }{E_{0}+\varepsilon (n_{\nu },l_{\nu },j_{\nu
})-\varepsilon (n_{\pi },l_{\pi },j_{\pi })}  \nonumber \\
&=&[n_{Z,N}^{J_{i}=0}n_{Z-2,N+2}^{J_{f}=0}]^{-1/2}\int\limits_{0}^{\pi
}n_{(Z,N),(Z-2,N+2)}(\theta ) \nonumber \\
&&\times \sum\limits_{\alpha \beta \gamma \delta }\frac{%
\left\langle \alpha \beta \left| \mathbf{\sigma }_{1}.\mathbf{\sigma }%
_{2}\tau ^{-}\tau ^{-}\right| \gamma \delta \right\rangle }{E_{0}+
\varepsilon (n_{\nu },l_{\nu },j_{\nu })-\varepsilon (n_{\pi },l_{\pi
},j_{\pi })}  \nonumber \\
&&\times \sum_{\varepsilon \eta }
\left[ 1+F_{Z,N}^{(\nu )}(\theta )f_{Z-2,N+2}^{(\nu )*}\right]^{-1}
_{\varepsilon \alpha } 
(f_{Z-2,N+2}^{(\nu )*})_{\varepsilon\beta }
\nonumber \\
&&\times \left[1+F_{Z,N}^{(\pi }(\theta )f_{Z-2,N+2}^{(\pi )*}\right]^{-1}_{\gamma \eta}
(F_{Z,N}^{(\pi )*})_{\eta \delta}
\sin\theta d\theta  \label{eqf}
\end{eqnarray}
where $\pi $ and $\nu $ refer to protons and neutrons in the parent and
daughter nuclei involved in the $\left( e^{+}\beta \beta \right) _{2\nu }$
decay modes.  Further,
\begin{equation}
\varepsilon (n_{\nu },l_{\nu },j_{\nu })-\varepsilon (n_{\pi },l_{\pi
},j_{\pi })=\left\{ 
\begin{array}{l}
\Delta _{C}-2E_{0}  \\ 
\Delta _{C}-2E_{0}+\Delta E_{s.o. splitting} 
\end{array}
\right.  \label{den}
\end{equation}
for  $n_{\nu }=n_{\pi },l_{\nu }=l_{\pi },j_{\nu}=j_{\pi }$ and 
$n_{\nu }=n_{\pi},l_{\nu }=l_{\pi },j_{\nu }\neq j_{\pi }$, respectively
and the Coulomb energy difference $\Delta _{C}$ is given by \cite{bohr98} 
\begin{equation}
\Delta _{C}=\frac{0.70}{A^{1/3}}\left[ \left( 2Z+1\right) -0.76\left\{
\left( Z+1\right) ^{4/3}-Z^{4/3}\right\} \right]
\end{equation}
The expressions to calculate $n^{J}$, $n_{(Z,N),(Z-2,N+2)}(\theta )$, $%
f_{Z-2,N+2}$\ \ and $F_{Z,N}(\theta )\ $are given in \cite{shuk05,rain06}.

In the present scheme, the difference in proton and neutron single particle
energies with the same quantum numbers corresponding to the energy of the
Isobaric Analog States is well described by the difference in Coulomb
energy $\Delta _{C}$. Further, the spin-orbit energy splitting is added
whenever required. The explicit inclusion of the spin-orbit splitting in the
energy denominator of equation (\ref{den}), implies that it cannot be factorized
out of the sum in equation (\ref{m2n}). In the present context, it is noteworthy
that the use of the summation method is richer than the closure
approximation, because each proton-neutron excitation is weighted depending
on its spin-flip or non-spin-flip character, and in addition, employing the
summation method in conjunction with the PHFB formalism goes beyond what was
done in previous applications of the pseudo-SU(3) model \cite{cero99,cast94,jghi95}.

\section{Results and discussions}

In the HFB framework, the nuclear wave functions for $^{156}$Dy and $^{156}$%
Gd isotopes are generated using the same model space and single particle
energies (SPE's), but for the SPE's of $0h_{11/2},$ $1f_{7/2}$ and $0h_{9/2}$
orbits, which are 4.6 MeV, 11.0 MeV and 11.6 MeV, respectively, as in our
earlier calculation on $\left( e^{+}\beta \beta \right) _{2\nu }$ decay
modes \cite{sing07}. The strengths of the pairing interaction are fixed as $%
G_{p}=G_{n}=30/A$ MeV. The strengths of the like particle components of the $%
QQ$ interaction are taken as $\chi _{pp}=\chi _{nn}=0.0105$ MeV $b^{-4}$,
where $b$ is oscillator parameter. For a given model space, SPE's, $G_{p}$, $%
G_{n}$ and $\chi _{pp}$, we fix the strength of proton-neutron ($pn$)
component of the $QQ$ interaction $\chi _{pn}$ by reproducing the excitation
energy $E_{2^{+}}$ of the 2$^{+}$ state. The adopted values of $\chi _{pn}$
(in MeV $b^{-4}$) for $^{156}$Dy and $^{156}$Gd isotopes in the present
calculation are 0.01817 and 0.02989, respectively. All these input parameters
are kept fixed throughout the subsequent calculations.

We present the calculated as well as experimentally observed results for
yrast spectra \cite{saka84}, reduced $B(E2$:$0^{+}\to 2^{+})$ transition 
probabilities \cite{rama01}, static quadrupole moments $Q(2^{+})$ and 
gyromagnetic factors $\ g(2^{+})$ \cite{ragh89} of $^{156}$Dy and $^{156}$Gd
nuclei in table~\ref{tab1}. We tabulate only the adopted value for the experimentally observed 
reduced $B(E2$:$0^{+}\to 2^{+})$ transition
probabilities \cite{rama01}. The calculated $B(E2$:$0^{+}\to 2^{+})$ are in excellent
agreement with the observed ones. No experimental $Q(2^{+})$ result is
available for $^{156}$Dy isotope. The agreement between the calculated and
experimental $Q(2^{+})$ for $^{156}$Gd nucleus is also quite good. However,
the calculated gyromagnetic factors $g(2^{+})$ for $^{156}$Dy and 
$^{156}$Gd isotopes are off from the experimental data.

In the calculation of $g$-factors for $^{156}$Dy and $^{156}$Gd isotopes, we use bare values
for $g_l^{\pi}=1$, $g_l^{\nu}=0$, $g_s^{\pi}$(effective)=0.5$g_s^{\pi}$(bare) and
$g_s^{\nu}$(effective)=0.5$g_s^{\nu}$(bare).
It is possible to include an effective way the role of higher $j$ orbitals not included 
in the model space, by using
a different set of effective gyromagnetic ratios, namely
 $g_l^{\pi}=0.7$,
$g_l^{\nu}=0.07$, $g_s^{\pi}$(effective)=0.5$g_s^{\pi}$(bare)
and $g_s^{\nu}$(effective)=0.8$g_s^{\nu}$(bare) as suggested by the analysis of
Rath and Sharma \cite{rath88}, which
can almost reproduce the $g$-factors $g(2^+)$ of all considered parent and daughter nuclei
in this model space. Specifically, the $g$-factors $g(2^{+})$
for $^{156}$Dy and $^{156}$Gd nuclei are 0.399 and 0.424 nm, respectively.

The phase space factors of $\left( \varepsilon \beta ^{+}\right) _{2\nu }$ and 
$\left(\varepsilon \varepsilon \right) _{2\nu }$ modes
for $^{156}$Dy are calculated following the notations of Doi \textit{et al} 
\cite{doi92} in the approximation $C_{1}=1.0,$ $C_{2}=0.0$, $C_{3}=0.0$ and $%
R_{1,1}(\varepsilon )=R_{+1}(\varepsilon )+R_{-1}(\varepsilon )=1.0$. The
calculated phase space factors are $G_{2\nu }\left( \varepsilon \beta
^{+}\right) =4.723\times 10^{-23}$ y$^{-1}$ and $G_{2\nu }\left(
\varepsilon \varepsilon \right) =2.962\times 10^{-20}$ y$^{-1}$. However,
it is more justified to use the nuclear matter value of $g_{A}$ around 1.0
in heavy nuclei. Hence, the theoretical half-lives $T_{1/2}^{2\nu }$ are calculated
for both $g_{A}=1.0$ and 1.261. In table~\ref{tab2}, we present the
calculated NTME $M_{2\nu }$ and half-lives $T_{1/2}^{2\nu }$ of 
$\left( \varepsilon \beta ^{+}\right) _{2\nu }$ and $\left(
\varepsilon \varepsilon \right) _{2\nu }$ modes for $^{156}$Dy
isotope along with other available theoretical results.
No experimental result is available for the $\left( e^{+}\beta \beta \right)
_{2\nu }$ decay modes of $^{156}$Dy isotope. Theoretically, the $0^{+}\rightarrow 0^{+}$ 
transition of $\left(\varepsilon \varepsilon \right) _{2\nu }$ mode for $^{156}$Dy 
has been investigated only in the
pseudo-SU(3) model \cite{cero99}. The calculated NTME $M_{2\nu }$ in the
PHFB model is 0.0138, which is smaller by a factor of 4.4 approximately than
that due to pseudo-SU(3) model \cite{cero99}. In the PHFB model, the
calculated half-lives $T_{1/2}^{2\nu }$ of $\left( \varepsilon \beta
^{+}\right) _{2\nu }$ and $\left( \varepsilon \varepsilon \right) _{2\nu }$
modes for $g_{A}=(1.261-1.0)$ are $(1.115-2.819)\times 10^{26}$ y and 
$(1.778-4.496)\times 10^{23}$ y, respectively.

The multipolar
correlations in general and quadrupolar correlations in particular present in 
the effective two-body interaction play a
crucial role in the deformation of nuclei. Therefore, 
a natural choice to
understand the role of deformation on the NTME $M_{2\nu }$ is to study its
variation with respect to changing strength of the $QQ$ interaction $%
\zeta_{qq}$. The results are presented in table~\ref{tab3} and figure~1(a).
\begin{figure}[tbh]
\begin{tabular}{cc}
\includegraphics [scale=0.26]{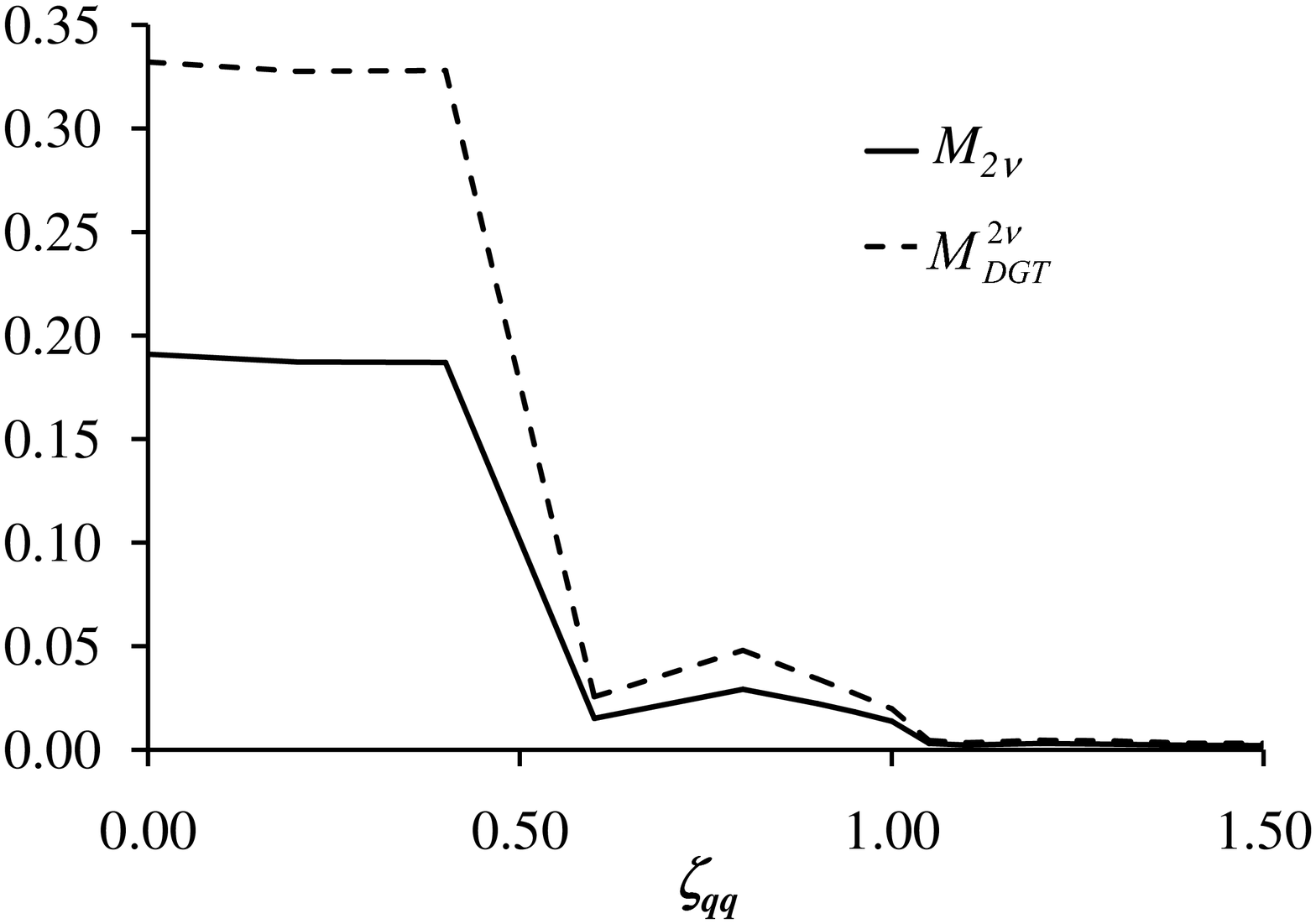} &
\includegraphics [scale=0.42]{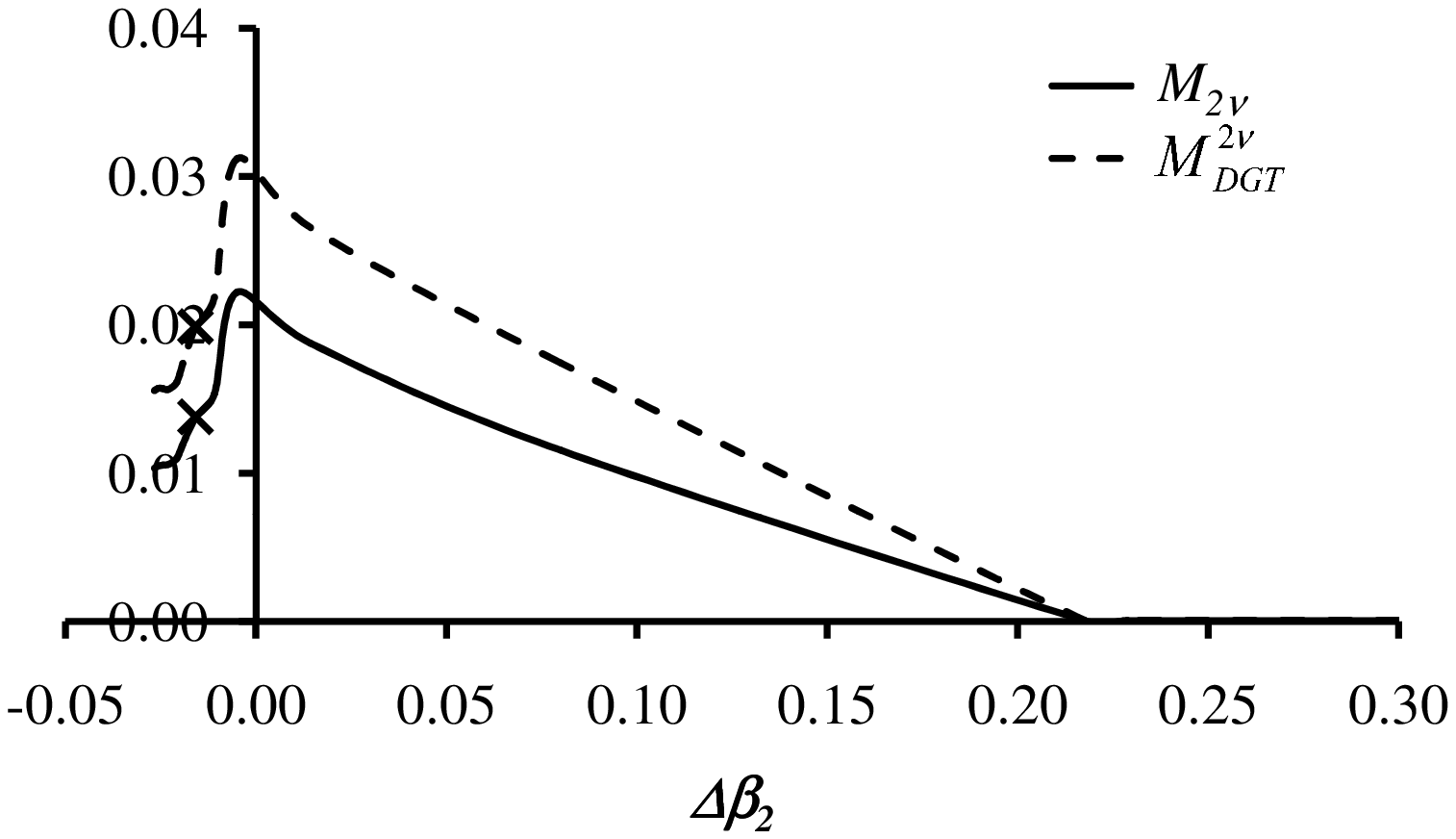}\\
(a)&(b)
\end{tabular}
\caption{(a) Dependence of $M_{2\nu }$ and $M_{DGT}^{2\nu}$ on the strength of 
$QQ$ interaction $\zeta _{qq}$.
(b) NTMEs $M_{2\nu }$ and DGT matrix elements  $M_{DGT}^{2\nu}$ of $\left( \varepsilon \beta
^{+}\right) _{2\nu }$ and $\left( \varepsilon \varepsilon \right) _{2\nu }$
modes for $^{156}$Dy nucleus as a function of the difference in the deformation
parameter $\Delta\beta_2$. ``$\times$" denotes the value of NTME for
calculated $\Delta\beta_2$ at $\zeta_{qq}=1$. The DGT matrix elements are 
scaled down by a factor of 20 in order to plot them
on the same scale as used for NTMEs calculated using the summation method.}
\label{fig1}
\end{figure}
It is observed that the $M_{2\nu }$ remains almost constant as the strength 
of $\zeta _{qq}$ is changed from $\zeta _{qq}=0.0-0.4$. As $\zeta _{qq}$ is
further increased up to 1.05, the NTME $M_{2\nu }$ starts decreasing except
at $\zeta _{qq}=0.80$, where a small increment in the size of $M_{2\nu }$ is
noticed. For a further variation of $\zeta _{qq}$ from 1.05 to 1.50, the
NTME $M_{2\nu }$ remains almost constant. In addition, a direct proportionality  
between the quadrupole moment $Q_{2}$ and deformation parameter $\beta _{2}$
exists. Specifically, the deformation parameter $\beta _{2}$ is
obtained using the expression given by Raman $et$ $al$ \cite{rama01}.
\begin{equation}
\beta _{2}=\left( \frac{4\pi }{3ZR_{0}^{2}}\right)\times \left(\frac{B(E2:0^{+}
\rightarrow 2^{+})}{e^{2}}\right)^{1/2}
\end{equation}
where
\begin{eqnarray}
R_{0}^{2} &=&(1.2\times 10^{-13}\,A^{1/3}\,cm)^{2} \nonumber \\
&=&0.0144\,A^{2/3}\,b 
\end{eqnarray}
In our earlier works \cite{rain06,sing07}, it has been shown
that the NTMEs $M_{2\nu }$ are usually large for $\zeta _{qq}=0.0$ i.e. when
both the parent and daughter nuclei are spherical. With the increase of $%
\zeta _{qq}$, the NTMEs remain almost constant and then decrease around the
physical value $\zeta _{qq}=1.0$ establishing an inverse correlation between 
$M_{2\nu }$ and deformation parameter $\beta _{2}$. Presently, a similar 
inverse correlation between the NTME $M_{2\nu }$ and $Q_{2}$ as well as 
$\beta _{2}$ also
exists. The effect of deformation on $M_{2\nu }$ is quantified by defining a
quantity $D_{2\nu }$ as the ratio of $M_{2\nu }$ at zero deformation $(\zeta
_{qq}=0)$ and full deformation $(\zeta _{qq}=1)$. The ratio $D_{2\nu }$
is 13.64 for $^{156}$Dy nuclei.

\begin{figure}[tbh]
\includegraphics [scale=1.23]{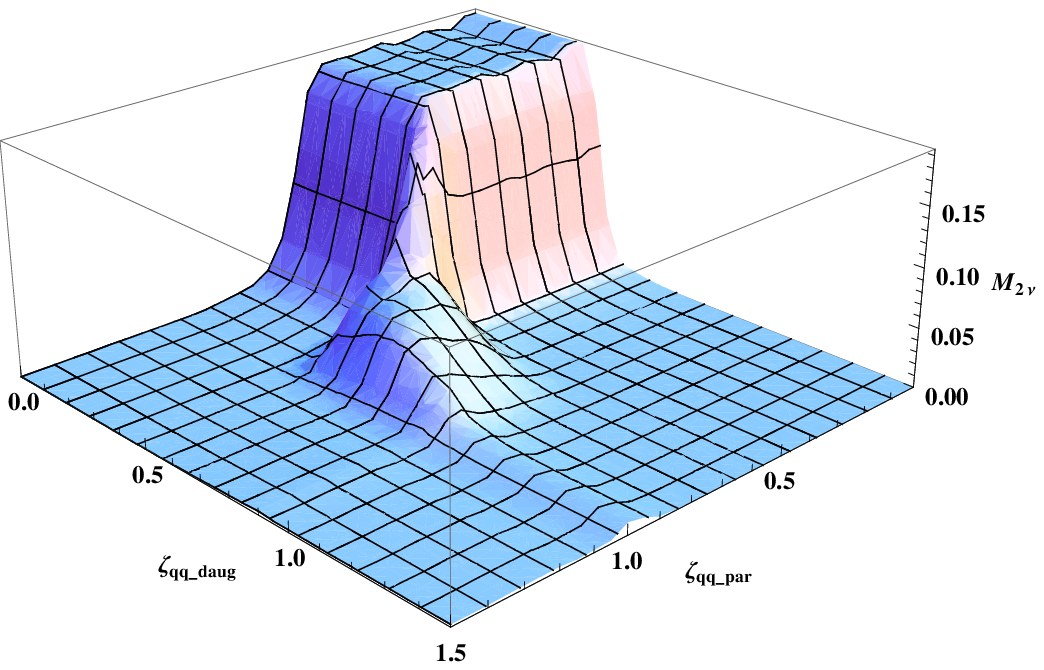}
\caption{Dependence of $M_{2\nu }$ on the independent variation of 
$\zeta _{qq}$ for parent and daughter nuclei.}
\label{fig2}
\end{figure}

To investigate the observed suppression of the NTMEs as observed in the  $\beta ^{-}\beta
^{-}$ decay with respect to the spherical case when the parent and daughter
nuclei have different deformations \cite{alva04,chan09,mene08}, we present
in figure~1(b), the NTME $M_{2\nu }$ as a function of the difference in the
deformation parameters $\Delta \beta _{2}=\beta _{2}(parent)-\beta
_{2}(daughter)$ of the parent and daughter nuclei.
The NTME is
calculated by keeping the deformation of parent nucleus fixed at $\zeta
_{qq}=1$ and the deformation of daughter nucleus is varied by taking $\zeta
_{qq}=0.0$ to 1.5. It is noticed from the figure~1(b) that the NTME is
maximum when the absolute of the difference in deformation $\left| \Delta
\beta _{2}\right| $ is minimum and when the $\left| \Delta \beta _{2}\right| 
$ increases the NTME $M_{2\nu }$ is strongly reduced. Thus, it is clear from
this behaviour of $M_{2\nu }$ with respect to changing $\zeta _{qq}$, that
the NTME tends to be large in the absence of quadrupolar correlations
i.e. for a pair of spherical $e^{+}\beta \beta $ parent and daughter $\beta\beta$ 
emitters. In
figure~\ref{fig2}, we present the 3-dimensional view of 
variation in NTME with respect
to $\zeta _{qq}=0.0-1.5$ for parent and daughter nuclei independently.
The plateau corresponds to the maximum value of $M_{2\nu }$ for small 
admixture of quadrupolar correlations corresponding to $\beta _{2}(parent)=0.0-0.086$ 
and $\beta_{2}(daughter)=0.0-0.163$. The next peak is due to the increase in 
the magnitude of $M_{2\nu }$ around $\zeta _{qq}=0.8$ and finally at $\zeta
_{qq}=1.0$, the small physical $M_{2\nu }$ is obtained.

In order to understand the connection between the closure approximation and the 
summation method, we extract the closure double Gamow-Teller (DGT) matrix element 
$M_{DGT}^{2\nu }$ and average energy denominator $E_d$. 
Further, the deformation effects due to the energy denominator are investigated by 
studying the variation 
of $M_{DGT}^{2\nu }$ with respect to $\zeta _{qq}$ and deformation parameters $\beta _{2}$ 
of parent and daughter nuclei. 
In the PHFB model, the extracted closure DGT matrix element $M_{DGT}^{2\nu }$ is 0.3975 and 
it turns out that the NTME 
$M_{2\nu }$ calculated in the summation method can be obtained by taking an 
average energy denominator $E_d=14.72$ MeV,
reasonably close to the value $E_d=13.99$ MeV recommended by Haxton and Stephenson
\cite{Hax84}. In Figs. 1(a) and 1(b), we also plot the variation of  closure 
$M_{DGT}^{2\nu }$ with respect to $\zeta _{qq}$ and $\Delta\beta _{2}$ respectively. 
It is clear from Figs. 1(a) and 1(b)  that 
the deformation effects on the closure DGT matrix elements and NTMEs calculated using 
the summation method are similar but not the same.  
\section{Conclusions}

To conclude, we have tested
the reliability of HFB intrinsic wave functions for $^{156}$Dy and $%
^{156}$Gd nuclei by calculating the spectroscopic properties, namely the
yrast spectra, reduced $B(E2$:$0^{+}\rightarrow 2^{+})$ transition
probabilities, static quadrupole moments $Q(2^{+})$ and $g$-factors $%
g(2^{+}) $ of these isotopes and comparing them with the available
experimental data. An overall agreement between the calculated and observed
spectroscopic properties suggests that the PHFB wave functions generated by
fixing $\chi _{pn}$ to reproduce the $E_{2^+}$ are quite reliable.
Subsequently, we employ the same PHFB wave functions to calculate NTME $%
M_{2\nu }$ as well as half-lives $T_{1/2}^{2\nu }(\varepsilon \beta ^{+})$ and 
$T_{1/2}^{2\nu }(\varepsilon \varepsilon )$ of $^{156}$Dy isotope for the 
$0^{+}\rightarrow 0^{+}$ transition. We also examine the effect of
deformation on NTME $M_{2\nu }$ by varying the strength of $QQ$ part of the
effective two-body interaction. It is noticed that the $M_{2\nu }$ is the largest 
in the absence of quarupolar correlations. Moreover, it is reduced by
a factor of 
13.6 
due to the quadrupolar
correlations, which may be taken as a conservative estimate of the
deformation effect in view of the approximations inherent in
the present calculation. Further, the NTME has a well defined maximum when the
deformations of parent and daughter nuclei are similar.
Employing the closure approximation for calculating the DGT matrix elements, it has been shown that the dependence on deformation is a general qualitative feature, which does not depend on details of the energy denominator.

This work was partially supported by DST, India vide sanction No.
SR/S2/HEP-13/2006, by Conacyt-M\'{e}xico and DGAPA-UNAM.

\begin{table}
\caption{Excitation energies (in MeV) of $J^{\pi }=2^{+},$ $4^{+}$
and $6^{+}$ yrast states, reduced $B(E2$:$%
0^{+}\rightarrow 2^{+})$  transition probabilities in $e^{2}$ b$^{2}$, 
static quadrupole moments \ $%
Q(2^{+})$ in $e$ b, $g$-factors $g(2^{+})$ in nuclear magneton for $^{156}$Dy
and $^{156}$Gd isotopes. Here $B(E2)$ and $Q(2^{+})$ are calculated for
effective charge $e_{p}=$1+$e_{eff}$ and $e_{n}=e_{eff}$ with $e_{eff}=0.60$
and $g(2^{+})$ has been calculated for $g_{l}^{\pi }=1.0$, $g_{l}^{\nu }=0.0$
and $g_{s}^{\pi }=g_{s}^{\nu }=0.50$.}
\label{tab1}
\begin{indented}
\item[]\begin{tabular}{llllll}
\hline\hline
& \multicolumn{2}{c}{$^{156}$Dy} && \multicolumn{2}{c}{$^{156}$Gd} \\ \cline{2-3} \cline{5-6}
& Theory & Experiment & &Theory & Experiment \\ \hline
&  &  & &  &  \\ 
$E_{2^{+}}$ & 0.1379 & 0.1378$^{a}$ && 0.0886 & 0.0889$^{a}$ \\ 
$E_{4^{+}}$ & 0.4572 & 0.4041$^{a}$ && 0.2936 & 0.2882$^{a}$ \\ 
$E_{6^{+}}$ & 0.9528 & 0.7703$^{a}$ && 0.6119 & 0.5847$^{a}$ \\ 
$B(E2)$ & 3.888 & 3.710$\pm $0.040$^{*b}$ && 4.052 & 4.64$\pm $0.05$^{*b}$ \\ 
$Q(2^{+})$ & -1.786 & - && -1.823 & -1.93$\pm $0.04$^{c}$ \\ 
&  &  &  & & -1.96$\pm $0.04$^{c}$ \\ 
$g(2^{+})$ & 0.550 & 0.39$\pm $0.04$^{c}$& & 0.602 & 0.387$\pm $0.004$^{c}$
\\ \hline\hline
\multicolumn{5}{l}{$^{a}${\small \cite{saka84}}; $^{b}${\small %
\cite{rama01}};$^{c}${\small \cite{ragh89}}} \\ 
\multicolumn{5}{l}{$^{\ast }${\small denotes the adopted value.}}
\end{tabular}
\end{indented}
\end{table}

\begin{table}
\caption{Theoretically calculated $M_{2\nu }$ and corresponding $%
T_{1/2}^{2\nu }(0^{+}\rightarrow 0^{+})$ of $\left( \varepsilon \beta
^{+}\right) _{2\nu }$ and $\left( \varepsilon \varepsilon \right) _{2\nu }$
modes for $^{156}$Dy nucleus. The $T_{1/2}^{2\nu}$
is calculated for $g_{A}=1.261$ and 1.0. * denotes the present
work.}
\label{tab2}
\begin{indented}
\item[]\begin{tabular}{llclll}
\hline\hline
Decay & Ref. & Model & $\left| M_{2\nu }\right| $ &\multicolumn{2}{c}{$T_{1/2}^{2\nu}$(y)}\\ \cline{5-6}
Mode &  & \multicolumn{1}{l}{} &  & $g_A=1.261$ & $g_A=1.0$ \\ \hline
&  & \multicolumn{1}{l}{} &  &  &  \\ 
$\varepsilon \beta ^{+}$ & * & \multicolumn{1}{l}{PHFB} & 0.0138 & 1.115$\times 10^{26}$ & 
2.819$\times 10^{26}$\\ 
&  & \multicolumn{1}{l}{} &  &  &  \\ 
$\varepsilon \varepsilon $ & * & \multicolumn{1}{l}{PHFB} & 0.0138 & 1.778$\times 10^{23}$ & 
4.496$\times 10^{23}$\\ 
& \cite{cero99} & \multicolumn{1}{l}{p-SU(3)} & 0.061 & 9.073$\times10^{21}$ & 2.294$\times 10^{22}$ \\ 
\hline\hline
\end{tabular}
\end{indented}
\end{table}

\begin{table}
\caption{Effect of the variation in $\zeta _{qq}$ on $%
\left\langle Q_{0}^{2}\right\rangle ,$ $\beta _{2}$ and $M_{2\nu }$ for $%
^{156}$Dy isotope.}
\label{tab3}
\begin{indented}
\item[]\begin{tabular}{ccccccc}
\hline\hline
$\zeta _{qq}$ & \multicolumn{2}{c}{$^{156}$Dy} && \multicolumn{2}{c}{$^{156}$%
Gd} & $\left|M_{2\nu }\right|$ \\ \cline{2-3} \cline{5-6}
& $\left\langle Q_{0}^{2}\right\rangle $ & $\beta _{2}$ && $\left\langle
Q_{0}^{2}\right\rangle $ & $\beta _{2}$ &  \\ \hline
\multicolumn{1}{l}{} & \multicolumn{1}{l}{} & &\multicolumn{1}{l}{} & 
\multicolumn{1}{l}{} & \multicolumn{1}{l}{} & \multicolumn{1}{l}{} \\ 
\multicolumn{1}{l}{0.00} & \multicolumn{1}{l}{0.000} & \multicolumn{1}{l}{
0.000} && \multicolumn{1}{l}{0.000} & \multicolumn{1}{l}{0.000} & 
\multicolumn{1}{l}{0.191} \\ 
\multicolumn{1}{l}{0.20} & \multicolumn{1}{l}{0.093} & \multicolumn{1}{l}{
0.073} && \multicolumn{1}{l}{0.071} & \multicolumn{1}{l}{0.071} & 
\multicolumn{1}{l}{0.187} \\ 
\multicolumn{1}{l}{0.40} & \multicolumn{1}{l}{0.345} & \multicolumn{1}{l}{
0.083} && \multicolumn{1}{l}{1.672} & \multicolumn{1}{l}{0.082} & 
\multicolumn{1}{l}{0.187} \\ 
\multicolumn{1}{l}{0.60} & \multicolumn{1}{l}{28.33} & \multicolumn{1}{l}{
0.102} && \multicolumn{1}{l}{65.81} & \multicolumn{1}{l}{0.220} & 
\multicolumn{1}{l}{0.015} \\ 
\multicolumn{1}{l}{0.80} & \multicolumn{1}{l}{65.98} & \multicolumn{1}{l}{
0.218} && \multicolumn{1}{l}{80.82} & \multicolumn{1}{l}{0.285} & 
\multicolumn{1}{l}{0.029} \\ 
\multicolumn{1}{l}{0.90} & \multicolumn{1}{l}{74.38} & \multicolumn{1}{l}{
0.250} && \multicolumn{1}{l}{86.39} & \multicolumn{1}{l}{0.305} & 
\multicolumn{1}{l}{0.022} \\ 
\multicolumn{1}{l}{0.95} & \multicolumn{1}{l}{82.82} & \multicolumn{1}{l}{
0.276} && \multicolumn{1}{l}{88.72} & \multicolumn{1}{l}{0.311} & 
\multicolumn{1}{l}{0.018} \\ 
\multicolumn{1}{l}{1.00} & \multicolumn{1}{l}{91.60} & \multicolumn{1}{l}{
0.300} && \multicolumn{1}{l}{90.78} & \multicolumn{1}{l}{0.316} & 
\multicolumn{1}{l}{0.014} \\ 
\multicolumn{1}{l}{1.05} & \multicolumn{1}{l}{94.27} & \multicolumn{1}{l}{
0.308} && \multicolumn{1}{l}{92.10} & \multicolumn{1}{l}{0.319} & 
\multicolumn{1}{l}{0.003} \\ 
\multicolumn{1}{l}{1.10} & \multicolumn{1}{l}{94.88} & \multicolumn{1}{l}{
0.310} && \multicolumn{1}{l}{92.58} & \multicolumn{1}{l}{0.321} & 
\multicolumn{1}{l}{0.002} \\ 
\multicolumn{1}{l}{1.20} & \multicolumn{1}{l}{95.42} & \multicolumn{1}{l}{
0.312} && \multicolumn{1}{l}{93.25} & \multicolumn{1}{l}{0.323} & 
\multicolumn{1}{l}{0.003} \\ 
\multicolumn{1}{l}{1.30} & \multicolumn{1}{l}{96.08} & \multicolumn{1}{l}{
0.315} && \multicolumn{1}{l}{93.66} & \multicolumn{1}{l}{0.324} & 
\multicolumn{1}{l}{0.003} \\ 
\multicolumn{1}{l}{1.40} & \multicolumn{1}{l}{96.62} & \multicolumn{1}{l}{
0.317} && \multicolumn{1}{l}{94.00} & \multicolumn{1}{l}{0.326} & 
\multicolumn{1}{l}{0.002} \\ 
\multicolumn{1}{l}{1.50} & \multicolumn{1}{l}{97.01} & \multicolumn{1}{l}{
0.318} && \multicolumn{1}{l}{94.32} & \multicolumn{1}{l}{0.327} & 
\multicolumn{1}{l}{0.002} \\ 
\hline\hline
\end{tabular}
\end{indented}
\end{table}
\end{document}